\documentstyle[12pt]{article}
\def\singlespacing{\baselineskip=12pt}

\begin{document}
\input{epsf}
\singlespacing

\title{\bf{\Large
{\bf Capacity of multivariate channels with multiplicative noise:
I.Random matrix techniques and large-N expansions for full
transfer matrices}}}
\author{
Anirvan Mayukh Sengupta\\
Partha Pratim Mitra\\
Bell Laboratories,  Lucent Technologies, Murray Hill, NJ 07974\\
}
\maketitle

\bigskip

\bigskip

\abstract{
We study memoryless, discrete time,
matrix channels with additive white Gaussian noise and input power
constraints of the form
$Y_i = \sum_j H_{ij} X_j + Z_i$, where $Y_i$ ,$X_j$ and $Z_i$
are complex, $i=1..m$, $j=1..n$, and $H$ is a complex $m\times n$ matrix
with some degree of randomness in its entries. The additive
Gaussian noise vector is assumed to have uncorrelated entries.
Let $H$ be a full matrix (non-sparse) with pairwise
correlations between matrix entries of the form
$ E[H_{ik} H^*_{jl}] = {1\over n}C_{ij} D_{kl} $, where $C$,$D$ are
positive
definite Hermitian matrices.
Simplicities arise in the limit of large matrix sizes (the so called
large-$n$ limit) which allow us to obtain several exact expressions
relating to the channel capacity.
We study the probability distribution of the quantity
$ f(H) = \log \det (1+P H^{\dagger}S H) $. $S$ is non-negative definite
and
hermitian, with $Tr S=n$.  Note that the
expectation $E[f(H)]$, maximised over $S$, gives the capacity of the
above
channel with an input power constraint in the case $H$
is known at the receiver but not at the transmitter.
For arbitrary $C$,$D$ exact expressions are obtained for the expectation
and variance of $f(H)$ in the large matrix size limit.
For $C=D=I$, where $I$ is the identity matrix,
expressions are in addition obtained for the full moment
generating function for arbitrary (finite) matrix size in the
large signal to noise limit. Finally, we obtain the channel
capacity where the channel matrix is partly known and partly unknown
and of the form $\alpha I+ \beta H$, $\alpha,\beta$ being known
constants and entries of $H$ i.i.d. Gaussian with variance $1/n$.
Channels of the form described above
are of interest for wireless transmission
with multiple antennae and receivers.
}

\bigskip

\newpage

\section{Introduction}

Channels with multiplicative noise are in general difficult to
treat and not many analytical results are known for the channel
capacity and optimal input distributions. We borrow techniques
from random matrix theory \cite{mehta}
and associated saddle point integration
methods in the large matrix size limit to obtain several analytical
results for the memoryless discrete-time matrix channel with
additive Gaussian noise. Apart from the intrinsic interest in
multiplicative noise, these results are relevant to the study
of wireless channels with multiple antennae and/or receivers
\cite{telatar95,foschini96,marzetta}.

The channel input-output relationship is defined as

\begin{equation}
Y_i = \sum_{j=1}^n H_{i j} X_j + Z_i
\end{equation}
where all the quantities are in general complex, and $i=1 ... m$,
$j=1 ... n$. $Z_i$ are Gaussian distributed with zero mean and
a unity covariance
matrix, $E[Z_i Z^*_j] = \delta_{ij}$. Note that this fixes
the units for measuring signal power. For most of the paper we
employ an overall power constraint

\begin{equation}
\sum_{j=1}^n E[|X_j|^2] = nP
\end{equation}
except in one case where we are able to employ an amplitude (or peak
power) constraint. The entries of the matrix $H_{ij}$ are assumed to
be chosen from a zero mean Gaussian distribution with covariance matrix

\begin{equation}
E[H_{ik} H^*_{jl}] = {1\over n}C_{ij} D_{kl}
\label{hcorr}
\end{equation}

Here $C,D$ are positive definite Hermitian matrices.
Note that although we assume the distribution of $H$ to be Gaussian,
this assumption can be somewhat relaxed without substantially affecting
some of the large $n$ results. This kind of universality is expected
from known results in random matrix theory \cite{mehta}. However, for
simplicity we do not enter into the related arguments.

We consider the case where $C,D$ are arbitrary
positive definite hermitian matrices,
as well as the special case where $C,D$ are identity matrices.
In either case, one needs
to consider the scale of $H$. Since $H$ multiplies $X$, we absorb the
scale
of $H$ into $P$. The formulae derived in the paper can be converted into
more explicit ones exhibiting the scale of $H$ (say $h$)
and the noise variance $\sigma$ by the simple substitution
$ P \rightarrow P h^2/\sigma^2 $.

A note about our choice of convention regarding scaling with $n$: We
chose to scale the elements of the matrix $H_{i j}$ to be order $
1/\sqrt{n}$
and let each signal element $X_j$ be order $1$. In 
the multi-antenna wireless  literature,
it is common to do the scaling the other way round.  In these 
papers \cite{telatar95, foschini96},  $X_j$'s are scaled  as
$1/\sqrt{n}$ but
keeping $H_{ij}$'s are kept order $1$ so that the average {\it total}
power
is $P$. Our choice of convention
is motivated by the fact that we want to treat the systems with channel
known at receiver and those with partially unknown
channel within the same framework.
For reasons that will become clear later, it is convenient for us to
keep the scaling of the input space and the output space to be the same,
i. e. to keep $Y_i$, $X_j$ and $Z_i$ all to be order $1$ and to scale
down
$H_{ij}$ to be order $1/\sqrt{n}$. The advantage of this is that the
singular values of $H$ happens to be order $1$. For the results in the
last section, it is convenient that the fluctuating part of the matrix
scales this way, in order to have a meaningful result . The final answer
for capacity is obviously the same in either convention.  While
using our results in the context of multiantenna wireless,   
we just have to remember
that the total power, in physical units, is $P$, and not $nP$.

In this paper, we discuss two classes of problems. The first class
consists
of cases where $H$ is known to the receiver but not to the
transmitter.$H$
being known to neither corresponds to problems of the second class.
The case where $H$ is known to both could be solved by a combination of
random matrix techniques used in this paper and the water-filling
solution
\cite{telatar95}.

 As for the first class of problems, we need to maximise the mutual
information
$I(X,(H,Y))$ over the probability distribution of $X$ subject to the
power
constraint. Following Telatar's argument \cite{telatar95}, one can show
that
it is enough to maximise over Gaussian distributions of $X$, with
$E(X)=0$.
Let $E(X_i^* X_j)=P S_{ij}$. $TrS=n$ so that the power constraint is
satisfied. $S$ has to be chosen so that $E(I(X,Y|H))$, i. e. mutual
information
of $X,Y$ for given $H$, averaged over different realisations of $H$, is
maximum.

Most of the paper deals with the statistical properties of
the quantity
\begin{equation}
f(H) = \log \det ( 1 + P H^{\dagger} S H) =
\sum_{i=1}^{rank(H)} \log(1+P \mu_i)
\end{equation}
where $\mu_i$ are the squares of the singular values of the matrix
$S^{1\over2}H$.

The conditions for optimisation over $S$ are as follows: Let
\begin{equation}
E(H ( 1 + P H^{\dagger} S H)^{-1}H^{\dagger}) =\Lambda
\label{optcond}
\end{equation}
$\Lambda$ is a nonnegative definite matrix. Then
\begin{itemize}
\item $S$ and $\Lambda$ are simultaneously diagonalizable.
\item In the simultaneously diagonalizing basis, let the diagonal
elements $S_{ii}=s_i$ and $\Lambda_{ii}=\lambda_i$. Then for all $i$,
such that $s_i>0$, $\lambda_i=\lambda$.
\item For $i$ such that $s_i=0$, $\lambda_i <\lambda$.
\end{itemize}
The derivation of these conditions are provided in Appendix A.

\section{Channel known at the receiver: arbitrary matrix size,
uncorrelated
entries}

We start with the simplest case, in which the matrix entries are i.i.d.
Gaussian, corresponding to $C=I,D=I$. In this case, one obtains
$S=I$ for the capacity achieving distribution \cite{telatar95}.
In this case, the joint probability density of the singular values of
$H$ is
explicitly known to be given by \cite{mehta}

\begin{equation}
P(\mu_1,\ldots,\mu_{min(m,n)}) = {1\over {\cal Z}}
\prod_{i< j}(\mu_i-\mu_j)^2
\prod_i \mu_i^{|m-n|} e^{-n\sum_i \mu_i}
\end{equation}
where the normalisation constant can be obtained as a consequence
of the Selberg integral formula (\cite{mehta}, Pg.354, Eq.17.6.5)

\begin{equation}
{\cal Z} = \prod_{j=1}^{min(n,m)} \Gamma(j) \Gamma(|m-n|+j)
\end{equation}
In the following, we assume (without loss of generality) $min(n,m) = n$.

This form has been utilised before to obtain the expectation of $f(H)$
in terms of integrals over Laguerre polynomials \cite{telatar95}.
However,
it is also fairly straightforward to obtain the full moment generating
function (and hence the probability density) of $f(H)$, particularly
at large $P$. Consider the moment generating function $F(\alpha)$ of the
random variable $f(H)$, given by

\begin{equation}
F(\alpha) = E[\exp(\alpha f(H)) ] = E[\prod_i (1+P \mu_i) ^{\alpha}]
\end{equation}

\subsection{Large P limit}

In the limit of large $P$, the expectation can be simply computed as an
application of the integral formula stated above. Note that the large
$P$ limit is obtained when $P$ is much larger than the inverse of the
typical smallest eigenvalue. For the case $m=n$, this would require
that $P>>n$, whereas if $m/n=\beta >1$, then we
require $P>>(\sqrt{\beta}-1)^{-1}$. Taking the large $P$ limit, we
obtain

\begin{equation}
F(\alpha) \approx (P)^{\alpha n} E[\prod_i \mu_i ^{\alpha}]
\end{equation}

\begin{equation}
E[\prod_i \mu_i ^{\alpha}] =
\prod_{j=1}^n {\Gamma(\alpha+|m-n|+j) \over \Gamma(|m-n|+j)}
\end{equation}
In this limit, it follows that

\begin{equation}
E[f(H)] \approx n \log(P) + \sum_{j=1}^n \psi(m-n+j) -n\log(n)
\end{equation}

\begin{equation}
V[f(H)] \approx \sum_{j=1}^n \psi^{\prime}(|m-n|+j)
\end{equation}
where $\psi(j) = \Gamma^{\prime}(j)/\Gamma(j)$. Setting $m/n = \beta$ and for
large n,
we get

\begin{equation}
E[f(H)] \approx n \log(\beta P/e)
\end{equation}
For $\beta>1$ and large $n$,

\begin{equation}
V[f(H)] \approx \log({m\over m-n}) = \log({\beta \over \beta-1})
\end{equation}
For $\beta=1$ and large $m(=n)$,

\begin{equation}
V[f(H)] \approx \log(m)+1+\gamma
\end{equation}
where $\gamma$ is the Euler-Mascheroni constant.

Laplace transforming the moment generating function, one obtains the
probability
density of ${\cal C} = f(H)$. In the large $P$ limit, the probability
density
is therefore given by $p({\cal C} - n \log(P/e))$ where $p(x)$ is given
by

\begin{equation}
p(x) = {1\over 2 \pi} \int_{-\infty}^{\infty} d\alpha
e^{-i\alpha n(\log(n)-1)-i x\alpha}
\prod_{j=1}^n {\Gamma(i\alpha+|m-n|+j) \over \Gamma(|m-n|+j)}
\end{equation}
An example of $p(x)$ is presented in Fig.1 for $m=n=4$.

\subsection{Arbitrary P}

For arbitrary $P$, $F(\alpha)$ does not simplify as above, but can
nevertheless
be written in terms of an $n\times n$ determinant as follows:

\begin{equation}
F(\alpha) = {\det M(\alpha) \over \det M(0)}
\end{equation}
where the entries of the complex matrix $M$ are given by
($i,j=1 ... n$)

\begin{equation}
M_{ij}(\alpha) = \int_0^{\infty}d\mu (1+P \mu)^{\alpha}
\mu^{i+j+|m-n|-2}
e^{-n\mu}
\end{equation}

To obtain this expression for $F(\alpha)$, one has to simply express the
quantity $\prod_{i\neq j} (\mu_i - \mu_j)$ as a Vandermonde determinant
and perform the integrals in the resultant sum.
The integral can be expressed in terms of a Whittaker function
(related to degenerate Hypergeometric functions),
and can be evaluated rapidly, so that for small values
of $m,n$ this provides a reasonable procedure for numerical
evaluation of the probability distribution of $f(H)$.

\section{Channel known at the receiver:
large matrix size, correlated entries.}

For the more general case of correlations between matrix entries as in
Eq.\ref{hcorr}, the matrix ensemble is no longer invariant under
rotations
of
$H$, so that the eigenvalue distribution used in the earlier section
is no longer valid. However, by using saddle point integration
\cite{sengupta95}, it is
still possible to compute the expectation and variance of $f(H)$ in the
limit of large matrix sizes. In this section, we simply state the
results for the expectation and variance, and explore the consequences
of the formulae obtained. The saddle point method used to obtain these
results was used in an earlier paper to obtain the singular value
density of random matrices \cite{sengupta95}
and is described in Appendix B .

The expectation and variance of $f(H)$ are given in terms of the
following
equations:

\begin{equation}
E[f(H)] =  \sum_{i=1}^m \log(w+\xi_i r)
+ \sum_{j=1}^n \log(w+ \eta_j q) - n q r - (m+n) \log(w)
\label{expectn}
\end{equation}

\begin{equation}
V[f(H)] = -2 \log|1-g(r,q)|
\end{equation}
where

\begin{equation}
w^2 = {1 \over P}
\end{equation}

\begin{equation}
g(r,q) =[{1\over n} \sum_{j=1}^n ({\eta_j \over w + \eta_j q})^2 ]
[{1\over n}\sum_{j=1}^m ({\xi_j \over w + \xi_j r})^2 ]
\end{equation}

In the above equations, $\xi, \eta$ denote the eigenvalues
of the matrices $\tilde{C}=S^{1\over2}CS^{1\over2},D$ respectively.
The numbers $r,q$ are determined
by the equations

\begin{equation}
r = {1 \over n}\sum_{j=1}^n {\eta_j \over w + \eta_j q}
\label{reqn}
\end{equation}

\begin{equation}
q = {1\over n}\sum_{j=1}^m {\xi_j \over w + \xi_j r}
\label{qeqn}
\end{equation}
These equations are expected to be valid in the limit of
large $m,n$ assuming that a sufficient number of the eigenvalues
$\xi, \eta$ remain nonzero. These equations could be used to
design optimal multi-antenna systems \cite{aris}.

\section{Calculating Capacity}
In this section we provide the step by step procedure for calculating
capacity
using the results from the previous sections.
We found that the optimal covariance matrix $S$ and the matrix $C$ could be
diagonalized together. Let us work in the diagonalizing basis.
Define $\tilde{C}$ as before. This is a diagonal matrix in this basis,
with diagonal elements $\xi_i=c_is_i$, where $c_i,s_i$ are the diagonal
elements of $C,S$ respectively. We assume that $c_i$'s are sorted in
decreasing order. That is, $c_1>c_2>\cdots>c_m$.
The optimality condition, Eq.\ref{optcond},
becomes:
\begin{equation}
{c_ir\over w+c_is_ir}=\lambda, \textrm{ for } i=1,...,p.
\label{covar}
\end{equation}
$p$ is the number for nonzero $s_i$'s. One way to see this is as
follows:
Take the expression in Eq.\ref{expectn}, replace $\xi$ by $c_is_i$
 and take its derivative with respect to non-zero $s_i$'s. Note that
$q,r$ changes a $\xi_i$ changes. However, this expression is
evaluated at a point which is stationary with respect to variation in
$q$
and $r$. Hence, to first order, changes of $q,r$ due to changes in $\xi$
do not have a contribution. We just change $\xi$ keeping $q,r$ fixed.
Since $\partial \xi_i/\partial s_i=c_i$, we got the expression in
Eq.\ref{covar}.

Eq.\ref{covar}, along with Eq.\ref{reqn} and Eq.\ref{qeqn},
 provide $p+2$ equations for $p+3$ unknowns, namely $r,q$ and
$s_i,i=1,..,p$.
The additional condition comes from total power constraint $\sum_i
s_i=P$.
Once we find such a solution, we could check whether the conditions
$s_i>0$ and
$\lambda_i=c_ir/w<\lambda$ is satisfied for all $i>p$. If any of
them is not satisfied,
we need to change $p$, the number of non-zero eigenvalues of $S$.
After getting a consistent set of solutions we  use Eq.\ref{expectn} to
calculate capacity.

Schematically, the algorithm is as follows:
\begin{enumerate}
\item Diagonalize $C$ and arrange eigenvalues in the decreasing order
along the diagonal.
\item Start with p=1.
\item Solve equations \ref{covar},\ref{reqn},\ref{qeqn} along with the
power constraint.
\item Check whether $s_i>0$ for $i=1,..,p$, and, $c_{p+1}r/w<\lambda$.
\item If any of the previous conditions are not satisfied, go back to
step $3$ with $p$ incremented by $1$. Otherwise, proceed to next step.
\item Calculate capacity using Eq.\ref{expectn}.
\end{enumerate}

\section{Channel known at the receiver: large matrix size, uncorrelated
entries}

The results of the previous section simplify if we assume that
the matrix entries are uncorrelated with unit variance. In this
case, the equations become

\begin{equation}
E[f(H)] =  m \log(w + r) + n \log(w + q) - n q r - (m+n) \log(w)
\label{capavg}
\end{equation}

\begin{equation}
V[f(H)] = -2 \log|1-{1 \over (w +  q)^2}
{ \beta \over (w + r)^2}|
\label{capvar}
\end{equation}

\begin{equation}
r = {1  \over w + q}
\end{equation}

\begin{equation}
q = {\beta \over w + r}
\end{equation}
First, consider the special case where $m=n$. In this case, we obtain

\begin{equation}
E[f(H)] =  n\big[ \log \Bigl({P \over e}\Bigr) +
 \log\Bigl(1+{1\over x}\Bigr) + {x\over P}\big]
\end{equation}

\begin{equation}
V[f(H)] = 2 \log\Bigl({(1+x)^2 \over (2 x+1)} \Bigr)
\end{equation}
where $x^2+x=P$ (x positive). For large $P$, the expectation and
variance
tend to $n \log(P/e)$ and $\log(P)$ respectively. Note that the variance
grows logarithmically with power, but does not depend on the number
of channels.

For $m,n$ not equal, one obtains expressions which are analogous
by solving the simultaneous equations above for $q$ and $r$ (which
lead to quadratic equations for either $q$ or $r$
by elimination of the other variable):

\begin{eqnarray}
r(w) & = & { - (w^2+m-n) + \Delta \over 2 w} \\
q(w) & = & { - (w^2-m+n) + \Delta \over 2 w} \\
\Delta & = & \sqrt{(w^2+m+n)^2 - 4 m n}
\end{eqnarray}
Substituting these formulae in Eq.\ref{capavg} and Eq.\ref{capvar}
gives the desired expressions for the expectation and variance
of the capacity $f(H)$.

\section{H unknown at both receiver, transmitter: large matrix size,
uncorrelated entries}

The case where $H$ is unknown both to the transmitter and receiver is
in general hard \cite{marzetta}.
For example, analytical formulae for the capacity
are not available even in the scalar case. However, in the case that
the matrix entries are uncorrelated, the problem reduces to an effective
scalar problem which exhibits simple behaviour at large m. To proceed,
one first obtains the conditional distribution $p(\vec{Y}|\vec{X})$.
This
can be done by noting that for fixed $\vec{X}$, $\vec{Y}$ is a linear
superposition of zero mean Gaussian variables and is itself Gaussian
with
zero mean and variance given by

\begin{equation}
E[Y_i Y^*_j] = (1+{1\over n}\sum_k |X_k|^2)\delta_{ij}
\end{equation}

Note that only the magnitude of the vector $\vec{X}$ enters into the
equation,
and the distribution of $\vec{Y}$ is isotropic. Effectively, since the
transfer
matrix is unknown both at the transmitter and receiver, only magnitude
information and no angular information can be transmitted. Since we are
free to choose the input distribution of $x=|\vec{X}|/\sqrt{n}$,
we can henceforth
regard $x$ as a positive scalar variable. As for $y=|\vec{Y}|/\sqrt{m}$
($\sqrt{m}$ is just to arrange the right scaling),we still have to
keep track of the phase space factor $y^{2m-1}$ which comes from
transforming
to $2m$ dimensional polar coordinates. Note that we need $2m$
dimensions
since $\vec{Y}$ is a complex vector. Thus, the problem can be treated as
if
it
were a scalar channel, keeping track only of the magnitudes $y$ and
$x$,
except that the measure for integration over $y$
should be $d\mu(y) = \Omega_{2m} y^{2m-1} dy$ where
$\Omega_{2m}$ is from
the angular integral. The conditional probability $p(y|x)$ is given by

\begin{equation}
p(y|x) = \left[{m\over \pi (1+x^2)}\right]^m \exp(-{m y^2 \over 2
(1+x^2)})
\end{equation}

The conditional entropy of $y$ given $x$ is easy to compute from the
original
obervation that the conditional distribution is Gaussian, and is
given by

\begin{equation}
H(y|x) = m E_x\left[\log \left( {\pi e\over m} (1+x^2)\right)\right]
\end{equation}
The entropy of the output is

\begin{equation}
H(y) =- E_x  \int d\mu(y) p(y|x) \log(E_{x^{\prime}} p(y|x^{\prime}))
\end{equation}
Thus, the mutual information between input and output is given by
subtracting the two expressions above and rearranging terms:

\begin{equation}
I = - E_x  \int d\mu(y) p(y|x) \log(E_{x^{\prime}}
[({1+x^2 \over 1+x^{\prime 2}} )^m \exp(-{m y^2 \over (1+x^{\prime
2})}+m)])
\end{equation}

The $y$ integral contains the factor
\begin{equation}
y^{2m-1} \exp( -{m y^2 \over (1+x^2)})
\end{equation}
which is sharply peaked around $y^2 = (1+x^2)$ for $m$ large.
Thus, the $y$ integral can be evaluated using Laplace's method to obtain
(for $m$ large)

\begin{equation}
I \approx -E_x \log E_{x^{\prime}}
[({1+x^2 \over 1+x^{\prime 2}} )^m \exp(-m {(1+x^2) \over (1+x^{\prime
2})}+m)]
\end{equation}

Applying Laplace's method again to perform the integral inside the
logarithm, assuming that the distribution over $x$ is given by a
continuous function $p(x)$, we finally obtain

\begin{equation}
I = {1\over 2} \log({2 m \over \pi}) + \int dx p(x)
\log [{x\over 1+x^2} {1\over p(x)}]
\end{equation}

The capacity and optimal input distribution is straightforwardly
obtained by
maximising the above. It is easier to treat the case where a
peak power constraint
is used, namely $x \leq \sqrt{P}$. In this case, the optimal input
distribution
is ($x \in [0,\sqrt{P}]$)

\begin{equation}
p(x) = {1\over \log(1+P)} {2 x \over 1+x^2}
\end{equation}
and the channel capacity is

\begin{equation}
{\cal C} = {1\over 2} \log({m \over 2 \pi}) + \log(\log(1+P))
\end{equation}

Notice that the capacity still grows with $m$, which is somewhat
surprising, but this growth is only logarithmic. Secondly, the
dependence on the peak power is through a double logarithm.

With an average power constraint
$ \int x^2 dx p(x) = P $
the optimal input distribution
is given by

\begin{equation}
p(x) = a {2  x \over 1+x^2} e^{-{x^2\over a (1+P) }}
\end{equation}
where $a$ is a constraint determined by the normalisation condition,
which yields the equation

\begin{equation}
a = \int_0^{\infty} {dy \over 1+y} e^{-{y\over a (1+P)}}
\end{equation}

The capacity is given by
\begin{equation}
{\cal C} = {1\over 2} \log({m \over 2 \pi}) + \log(a) +{P\over 1+P}
{1\over
a}
\end{equation}
For large $P$, $a \approx \log(1+P)$, thus recovering the double
logarithm
behaviour.

\section{Information loss due to multiplicative noise}

We could generalize the calculation in the previous section to a problem
which interpolates
smoothly between usual additive noise channel and the case considered
above.
This is a problem with same number of transmitters and receivers
($m=n$) and is defined by
\begin{equation}
Y_i = \sum_{j=1}^n (\alpha \delta_{ij}+\beta H_{i j}) X_j + Z_i
\end{equation}
$\beta=0$ is the usual channel with additive gaussian noise.
$\alpha=0$ corresponds the problem we have just discussed. In the first
case,
capacity increases logarithmically with input power, whereas in the
second case
it has a much slower (double logarithmic) dependence on input power.
Apart
from the theoretical interest in studying the crossover between these
two kinds of behavior, this problem has much practical importance
\cite{jason}.

The easy thing to calculate is  $c=\lim_{n \rightarrow \infty} {\cal
C}/n$.
Notice that this quantity is zero in the limit $\alpha \rightarrow 0 $,
capacity
being logarithmic in $n$ in that limit. For simplicity, we choose the
input power constraint $ \sum_i |X_i|^2 \leq n P $. We relegate the
details
of the saddle point calculation to Appendix C. The result is

\begin{equation}
c=\log \left[1+{\alpha^2 P \over 1 + \beta^2 P}\right]
\end{equation}
The result tells us that, in the large $N$ limit,
 the effect of multiplicative noise is similar to
that if an additive noise whose strength increases with the input power.

It is of particular interest to note that there exists a lower bound
to the channel capacity, which is given by the capacity of a fictitious
additive gaussian channel with the same covariance matrix for
$(\vec{X},\vec{Y})$ as the channel in question.
Remarkably, this bound coincides with the saddle point answer.

\section{Appendix A}
The condition of optimality with respect to $S$ is
\begin{equation}
E[Tr\{(1+PH^{\dagger}SH)^{-1}H^{\dagger}\delta SH\}]
=Tr(\Lambda\delta S)\le 0
\end{equation}
for all allowed small $\delta S$. $\delta S$ has to satisfy two
conditions:
that $S+\delta S$ is non-negative definite and that $Tr(\delta S)=0$.
The matrix $\Lambda$ has been defined in the first section. It is a
non-negative definite  hermitian matrix.

If $S$ has only positive
eigenvalues then adding a small enough hermitian $\delta S$ to it
does not make any of the eigenvalues zero or negative. Then only way
the optimisation condition can be satisfied is by choosing $\Lambda$
to be proportional to the Identity matrix. This can be seen as follows:
for $\Lambda=\lambda I, Tr\Lambda \delta S=\lambda Tr\delta S=0$.
If $\Lambda \neq \lambda I$, then, in general, $Tr\Lambda \delta S\neq
0$
even though $\delta S=0$, and can therefore be chosen to be positive.

What if $S$ has  few zero eigenvalues?
Let us choose a basis so that $S$ is diagonal. The eigenvalue of  $S$
$s_i$ are ordered so that $s_1,\ldots,s_k$ are positive and
$s_i=0$ for $i>k$. We could choose $\delta S_{ij}$ to be non zero only
for
$1\le i,j \le k$  and repeating the argument of the last
paragraph, $\Lambda_{ij}=\lambda \delta_{ij}$, for $1\le i,j \le k$.
In fact, even if we choose $\delta S_{ij}$ to be nonzero for $i\le k
<j$, and
$j\le k <i$ we do not violate, to first order in $\delta S$, non
negativity
of eigenvalues of $S+\delta S$. This would give us $\Lambda_{ij}=0$ for
$i\le k <j$ and $j\le k <i$. Hence $\Lambda$ is of block-diagonal form.
The $k \times k$ block is already constrained to be proportional
to Identity matrix. We would now constrain the other block of $\Lambda$
which is of size $(n-k)\times (n-k)$.

Since the last $n-k$ eigenvectors of $S$ correspond to zero eigenvalues,
we are free to rotate them among each other. Using this freedom, we
diagonalise the lower $(n-k)\times (n-k)$ block of $\Lambda$. Choosing
diagonal $\delta S_{ij}$ with with negative values for $i=j\le k$
but positive values $i=j>k$, and satisfying $Tr(\delta S)=0$, we
can show that the last $n-k$ eigenvalues of $\Lambda$ are smaller than
or equal to $\lambda$.

\section{Appendix B}

In this section, it is assumed without loss of generality that $m \geq
n$.
We consider first the case $S=I$, but derive the results for arbitrary
$C,D$.
It is easy to recover the results for general $S$ by making the
transformation
$H\rightarrow S^{1\over 2}H$ and $C\rightarrow S^{1\over 2} C S^{1
\over 2}$.

We start from the identity
\begin{equation}
\det([w~ iH~; ~-iH^{\dagger} ~w])^{-\alpha} =
\int d\mu(X) d\mu(Y) \exp(-{1\over 2}
\sum_{a=1}^{\alpha} [ w (Y^{\dagger}_a Y_a + X^{\dagger}_a X_a) \\
+i(Y^{\dagger}_a H X_a - X^{\dagger}_a H^{\dagger} Y_a)])
\end{equation}
where

\begin{equation}
d\mu(X) = \prod_{i=1}^n \prod_{a=1}^{\alpha}
{ dX^R_{ia} dX^I_{ia} \over 2\pi}
\end{equation}
with $R,I$ denoting real and imaginary parts respectively.
$d\mu(Y)$ is defined analogously. The introduction of multiple
copies of the Gaussian integration is the well known `replica trick'.
This allows us to compute $f(H)$, since it is easily verified that

\begin{equation}
\det([w~ iH~; ~-iH^{\dagger} ~w])^{-\alpha}=w^{-(m+n)\alpha}
e^{-\alpha f(H)}
\end{equation}
where we have set $w^2=n/P$.
The moment generating function of $f(H)$ can be obtained
by studying the expectation of the determinant above with respect to
the probability distribution of $H$. We therefore obtain for the
moment generating function

\begin{equation}
F(-\alpha) = w^{(m+n)\alpha} \int d\mu(X) d\mu(Y) \exp(-{1\over 2}
[ w \sum_{a=1}^\alpha (Y^{\dagger}_{a} Y_a + X^{\dagger}_{a} X_a)\\
+{1\over 2n}\sum_{a,b=1}^\alpha
( Y^{\dagger}_{a} C Y_{b}  X^{\dagger}_{b} D X_{a})])
\end{equation}

The last term in the exponent can be decoupled by introducing
the $\alpha\times \alpha$ complex matrices $P,Q$ with contour integrals
over the matrix entries in the complex plane to obtain

\begin{equation}
F(-\alpha) = w^{(m+n)\alpha} \int d\mu(X) d\mu(Y) d\mu(R) d\mu(Q)
 \exp(-{1\over 2} S)
\end{equation}
where

\begin{equation} S =
 w \sum_{a=1}^\alpha (Y^{\dagger}_{a} Y + X^{\dagger}_{a} X)
+\sum_{a,b=1}^\alpha
( Y^{\dagger}_{a} C Y_{b}  R_{a b}
+  Q_{a b}  X^{\dagger}_{a} D X_{b}
- nR_{a b} Q_{b a} )
\end{equation}

\begin{equation}
d\mu(R) d\mu(Q) = \prod_{a,b=1}^{\alpha} {dR_{a b} dQ_{ab} \over
2\pi}
\end{equation}

The $R,Q$ integrals, in contrast with the $X,Y$ integrals, are complex
integrals along appropriate contours in the complex plain. For example,
if the $Q_{ij}$ integrals are along the imaginary
axis, so that the $Q$ integrals give rise to delta functions which
can then be integrated over $R$ to obtain the above equation.
The integrals over $X,Y$ can now be performed to obtain

\begin{equation}
F(-\alpha) = w^{(m+n)\alpha} \int d\mu(R) d\mu(Q)
\exp(-\log(\det(w+C R)) -\log(\det(w+D Q)) + nTr (R Q))
\end{equation}
where $CR$ and $DQ$ are understood to be outer products of the
matrices. Introducing the eigenvalues $\xi, \eta$ of $C,D$ the
exponent may be written as

\begin{equation}
\sum_{i=1}^m \log(\det(w+\xi_i R) )
+ \sum_{j=1}^n \log(\det(w+ \eta_j Q)) -n Tr(RQ)
\end{equation}

If $m,n$ become large and the number of non-zero
$\xi_i, \eta_i$ grow linearly with $m,n$, then we
can perform the $R,Q$ integrals using saddle point methods.
If we assume that at the saddle point the matrices $R,Q$
do not break the replica symmetry , i.e $R = r I$, $Q = q I$
where $I$ is the identity matrix, then the saddle point equations
are $\partial {\cal C} /\partial r = \partial {\cal C} /\partial q = 0$,
where ${\cal C}$ is defined below, leading to

\begin{equation}
r = {1\over n}\sum_{j=1}^n {\eta_j \over w + \eta_j q}
\end{equation}

\begin{equation}
q ={1\over n} \sum_{j=1}^m {\xi_j \over w + \xi_j r}
\end{equation}

Expanding the exponent upto quadratic order around the saddle
point and performing the resulting Gaussian integral, we
obtain

\begin{equation}
F(\alpha) = \exp(\alpha {\cal C}(r,q) + {\alpha^2\over 2} {\cal V}(r,q))
\end{equation}

\begin{equation}
{\cal C}(r,q) = \sum_{i=1}^m \log(w+\xi_i r)
+ \sum_{j=1}^n \log(w+ \eta_j q) - nq r - (m+n) \log(w)
\end{equation}

\begin{equation}
{\cal V}(r,q) = -2 \log|1-g(r,q)|
\end{equation}

\begin{equation}
g(r,q) =[{1\over n} \sum_{j=1}^n ({\eta_j \over w + \eta_j q})^2 ]
[{1\over n}\sum_{j=1}^m ({\xi_j \over w + \xi_j r})^2 ]
\end{equation}
Since $F(\alpha)$ is the moment generating function for $f(H)$,
the expressions for ${\cal C,V}$ give the expressions for the
expectation and variance of $f(H)$, as presented in section (3).

\section{Appendix C}
In this case,
\begin{equation}
P(\vec{Y}|\vec{X})={1 \over [ \pi (1+\beta^2 |X|^2)]^n}
           e^{-{|\vec{Y}-\alpha\vec{X}|^2 \over  (1+\beta^2 |X|^2/n)}}
\end{equation}
Let us redefine $\vec{x}=\vec{X}$ and $\vec{y}=\vec{Y}/\sqrt{n}$.
The optimal probability distribution of $\vec{x}$ depends only
on its norm $x=|\vec{x}|/\sqrt{n}$. Let $q(x)$ to be the
probability distribution
of $x$.

Once more,
\begin{equation}
H(\vec{y}|\vec{x})=E_{\vec{x}}\left[n \log\left( \pi e (1+\beta^2
x^2)/n\right)
                                              \right]
                  =n\int dx q(x)\log\left[{ \pi e \over n}(1+\beta^2
x^2)
\right]
\end{equation}
However,
\begin{equation}
p(\vec{y})=E_{\vec{x}}\left[p(\vec{y}|\vec{x})\right]
\approx \int dx q(x) {n^n \over[ \pi (1+\beta^2 x^2)]^n}
    e^{-{n(y^2+\alpha^2 x^2)\over (1+\beta^2 x^2)}+2n\phi\left({\alpha
xy\over
                  1+\beta^2 x^2}\right)}
\end{equation}
where
\begin{equation}
\phi(a)=\lim_{d\rightarrow \infty} {1 \over d}
\log\left[{\int_0^{\pi}d\theta \sin^{d-2}(\theta)e^{da\cos(\theta)}
\over \int_0^{\pi}d\theta \sin^{d-2}(\theta)} \right]
\end{equation}

Saddle point evaluation of $\phi(a)$ (which is equivalent to doing
an expansion of the Bessel functions $I_\nu(z)$ with large order $\nu$
and large argument $z$, but the ratio $z/\nu$ held fixed) gives

\begin{eqnarray}
\phi(a)&=&a \cos\theta(a)+\log\sin\theta(a) \\
  \cos\theta(a)&=& a \sin^2\theta(a)
\end{eqnarray}
In fact we would need $d\phi(a)/da$.
\begin{equation}
{d\phi(a) \over da}=\cos\theta(a)={\sqrt{1+4a^2}-1 \over 2a}
\end{equation}
Variation of $H(\vec{y})=\int d\vec{y}p(\vec{y})\log{1 \over
p(\vec{y})}$
with respect to $q(x)$ produces
\begin{equation}\label{deriv}
{\delta H(\vec{y})\over \delta q(x)}
=-\int d\vec{y}p(\vec{y}|x)(1+\log p(\vec{y}))
\end{equation}
where
\begin{equation}
p(\vec{y}|x)=\left[{n \over  \pi (1+\beta^2 x^2)}\right]^n
\exp(-nf(x,y))
=p(y|x)
\end{equation}
and
\begin{equation}
f(y,x)={y^2+\alpha^2 x^2 \over (1 +\beta^2 x^2)}
               -2 \phi({\alpha x y \over 1+\beta^2 x^2})
\end{equation}

Now we can do the $\vec{y}$ integral in Eq.\ref{deriv} by the
saddle point method.
After going over to polar coordinates and doing some straightforward
calculations, we find that the integral peaks
at $y=y(x)$ given by
\begin{equation}\label{peak}
y(x)^2=(1+(\alpha^2+\beta^2)x^2)
\end{equation}
This is expected, as variance of $\vec{y}$ given a uniform angular
distribution of $\vec{x}$  with a fixed norm $x$ is the right hand side
of
(\ref{peak}). On the other hand, the variance is $y(x)^2$ in the saddle
point
approximation.

Thus finally, we have the condition for the stationarity of the
mutual information,
\begin{equation}
-{\cal C}=\log\int dx'q(x')p(y(x)|x')
                         +n\log\left[{\pi e\over n}(1+\beta^2
x^2)\right]
\end{equation}
where ${\cal C}$ is a constant, which turns out to be the channel
capacity.
The constant is fixed by the condition that $q(x)$ is a normalised
probability distribution. This condition, along with the fact
$\int d\vec{y}p(y|x)=\Omega_{2n}\int dy y^{2n-1}p(y|x)=1$,
$\Omega_{2n}=2\pi^n/\Gamma(n)$, can be used to determine $C$.
\begin{eqnarray}
1&=&\Omega_{2n}\int dx y'(x) y(x)^{2n-1}\int dx'q(x')p(y(x)|x') \\
&=&e^{-{\cal C}}\Omega_{2n}\int_0^{\sqrt{P}}dx
    \left[{n\over \pi e(1+\beta^2 x^2)}\right]^n {y'(x)\over
y(x)}y(x)^{2n}
\\
&\approx& e^{-{\cal C}} \sqrt{2n \over \pi} \int_0^{\sqrt{P}}dx
{y'(x) \over y(x)} \left[{y(x)^2 \over( 1+\beta^2 x^2)}\right]^n
\end{eqnarray}

For any $\alpha>0$,
\begin{equation}
f(x)=\log\left[{y(x)^2 \over( 1+\beta^2 x^2)}\right]
=\log\left[{1+(\alpha^2+\beta^2)x^2
\over 1+\beta^2 x^2}\right]
\end{equation}
 is a monotonically increasing function of $x$, for positive $x$. Hence
the last integral is dominated by the contribution from the region near
the
upper limit. For a monotonically increasing function $f(x)$,
\begin{equation}
\int_0^{z}g(x)\exp(nf(x))\approx {g(z)\exp(nf(z))\over nf'(z)}.
\end{equation}
Using this, we get
\begin{equation}
c=\lim_{n\rightarrow \infty}{\cal C}/n=\log\left[{1 +(\alpha^2+\beta^2)
P
\over 1+\beta^2 P}\right]
\end{equation}

{\bf Acknowledgements:} The authors would like to thank Emre
Telatar for many useful and inspiring discussions.
\newpage
\noindent{\bf Figure Captions}

\begin{description}

\item [Figure 1.]
The probability density function of $f(H)$ is given for
$m=n=4$ in the limit of large $P$. The origin is shifted
to the value $4\log(P/e)$.

\end{description}


\end{document}